\documentclass{kapproc} 

\usepackage{procps} 

\usepackage[dvips]{graphicx}

\upperandlowercase

\kluwerbib 

\begin{document}
\articletitle[Star-forming galaxies in the 2dFGRS]
{Star-forming galaxies in the 2dFGRS}
\author{Birgit Kelm, Gennaro Sorrentino \& Paola Focardi}
\affil{Dipartmento di Astronomia, Universit\'a di Bologna }
\anxx{Kelm\, Birgit}
\begin{abstract}
We examine the environment of star-forming galaxies selected in the 
2dFGRS. We find that bright star-forming galaxies display a significant 
deficit of faint neighbours relative to passive galaxies. 
If a deficit in fainter companions implies a smaller mass halo, 
data support a scenario predicting star-formation to be 
suppressed in all systems more massive than 10$^{13}$M$_{\odot}$, 
rather than in clusters only.  
\end{abstract}
\section{Introduction}
Data from the 2dFGRS (Lewis et al. 2002) and the SDSS (Gomez et al. 2003, 
Hogg et al. 2003) indicate that there is a deficit of star-forming galaxies in dense regions. 
Direct evidence of star-formation suppression in regions less extreme 
than rich clusters remains, however, controversial   
(Carlberg et al. 2001, Bower \& Balogh 2003).   
To explore whether the suppression of star-forming activity 
and the local number of neighbours are related, we analyse the environment 
of  star-forming and passive galaxies in the 2dFGRS 
(type 3+4 and type 1 respectively as in Madgwick et al. 2002).   
The sample  includes all 10695 galaxies in the redshift range [0.0005-0.25] 
with $b_j$ between 17 and 17.5. For each galaxy we have automatically 
identified neighbours, i.e. galaxies lying within an area of 1$h^{-1}$Mpc 
radius (computed at the redshift of the galaxy) and 
$\Delta$cz\,$\pm$1000\,km\,s$^{-1}$. 
Because the 2dFGRS is complete for b$_j$\,$\in$\,[17 - 19.5]  
we identify all neighbours from equally-luminous to $\sim$2-mag-fainter. 
\section{Faint and equally luminous neighbours}
The average number of neighbours of 2dF galaxies, 
averaged in bins of absolute magnitude, is shown in Fig.1. 
The average for all galaxies (empty circles) and for passive (squares) 
and star-forming (triangles) subsamples are plotted. The left panel shows 
the average when counting neighbours whose luminosity is $\approx$\,equal 
($|$$\Delta$b$_j$$|$\,$\leq$0.5) to the one of the galaxy itself 
while the right one shows the average when counting much fainter neighbours 
(1.5$\leq$\,$\Delta$b$_j$\,$\leq$2.5).  
Fig. 1 shows that star-forming galaxies inhabit regions which are less  
dense than passive galaxies. 
However, while fainter galaxies segregate in both panels, bright 
galaxies only segregate when faint neighbours are counted.  
The right panel indicates that among the brightest galaxies (Mag$\leq$-20), 
only passive galaxies display an increasingly excess of faint neighbours. 
The difference in the number of faint neighbours might trace a difference 
in the mass of the halos hosting bright passive and bright star-forming 
galaxies (Magliocchetti \& Porciani 2003).   
Star-forming galaxies might be in a small mass halos where the 
cooling gas is channeled to one or two galaxies, while passive galaxies 
might be in more massive halos (M$>$10$^{13}$M$_{\odot}$) which 
are accreating smaller ones, and whose pre-existing galaxies survive as 
distinct entities(Berlind et al. 2003). 
To assess that the lack of faint neighbours around bright star-forming 
galaxies indicates that they reside in smaller mass systems also 
matches observations indicating that diffuse X-emission 
is restricted to groups with a dominant E-S0 (Mulchaey et al. 2003, 
Helsdon et al. 2001). 
Observations are thus supporting the hyphothesis that 
star-formation is suppressed in bright galaxies residing in any 
group more massive than 10$^{13}$M$_{\odot}$, rather than in clusters only.  
\begin{figure}
{\includegraphics[width=2.in]{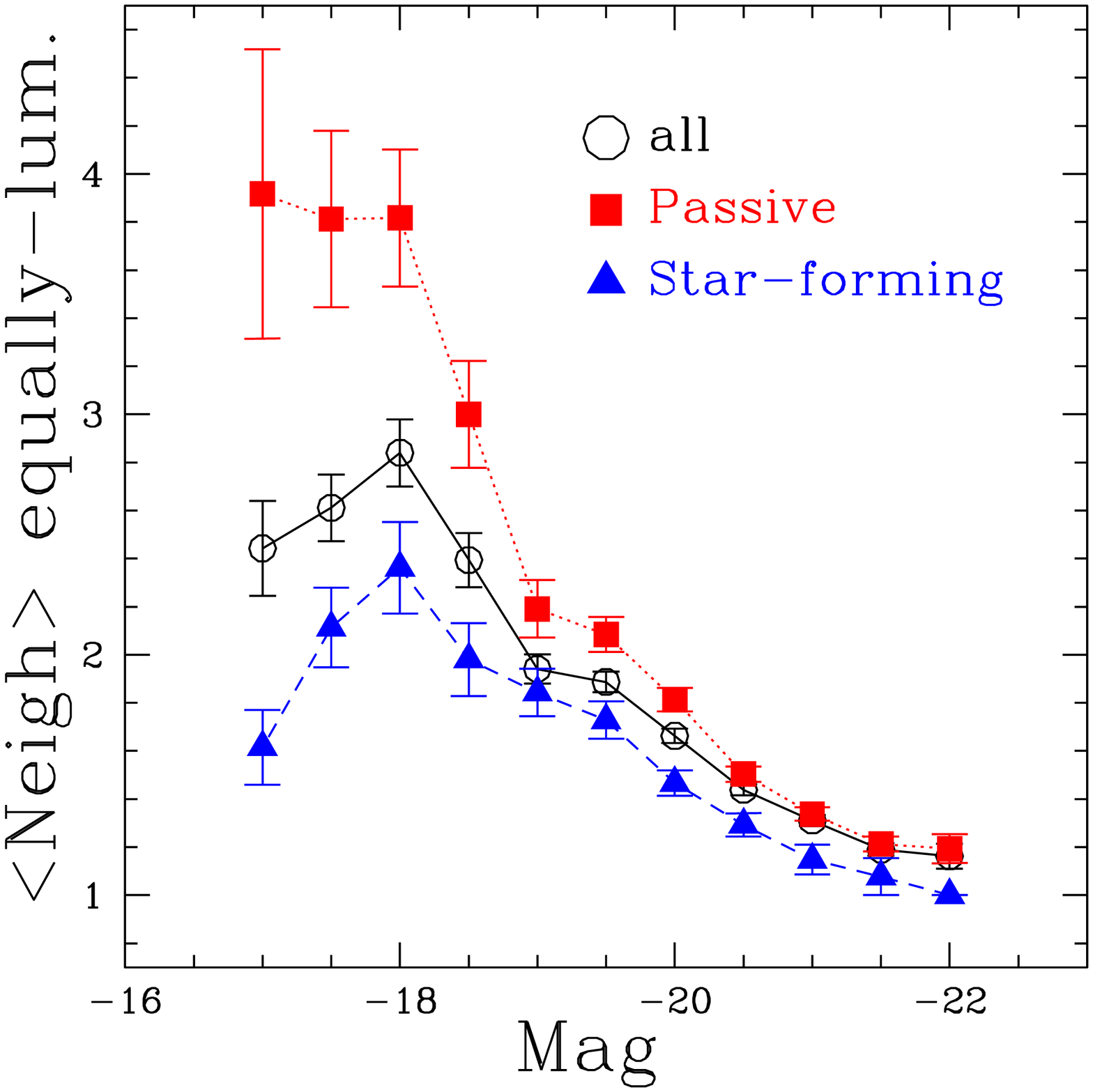}} 
{\includegraphics[width=2.in]{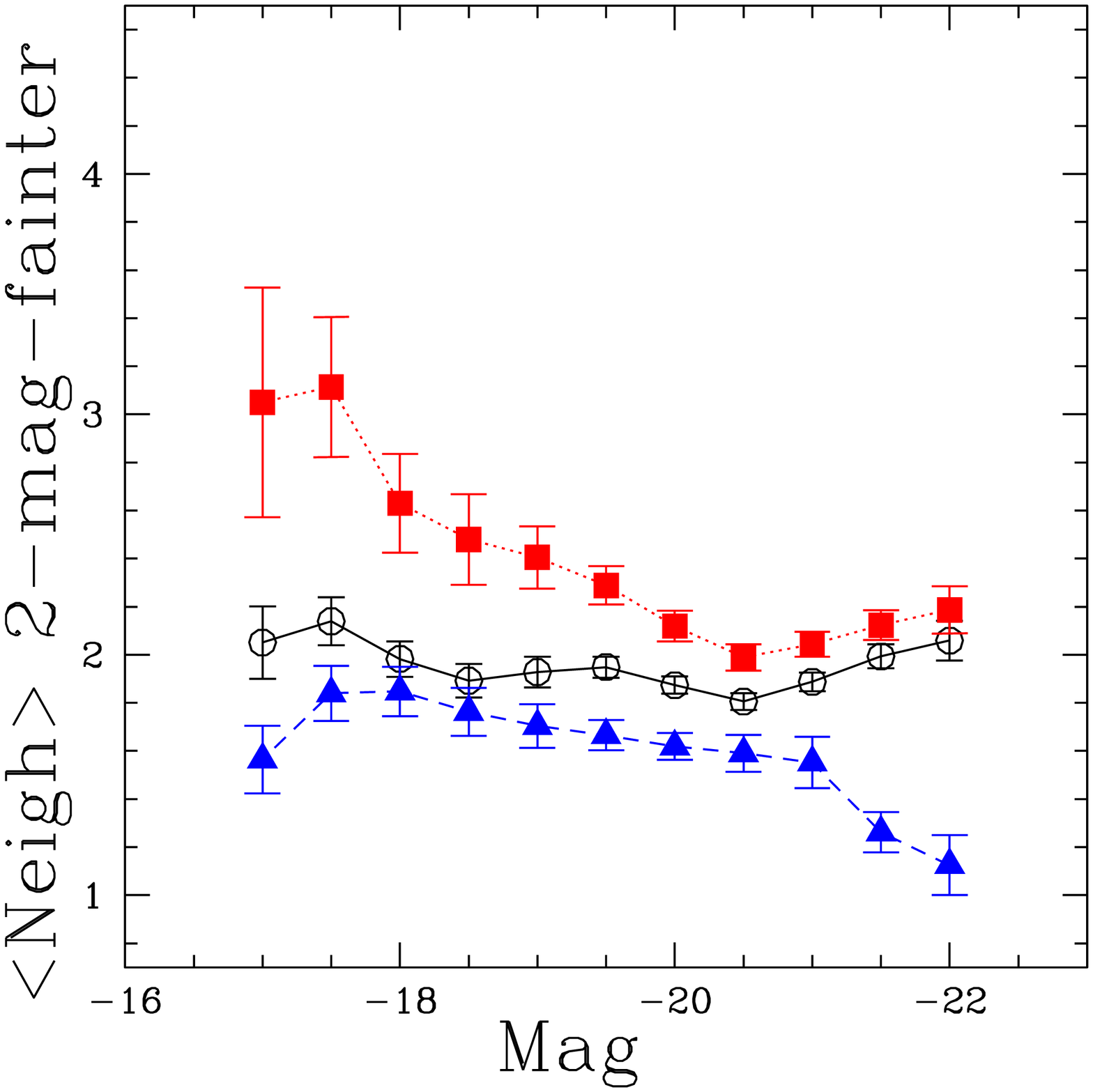}} 
\caption{Average number of equally-luminous (left) and fainter (right) neighbours}
\end{figure}

\end{document}